\documentclass[12pt,letterpaper]{article}
\usepackage[utf8]{inputenc}
\usepackage[margin=1in]{geometry}
\usepackage[english]{babel}
\usepackage{geometry}
\usepackage{dcolumn}
\usepackage{float}
\usepackage{rotating} % Add this in your preamble
\usepackage{comment}
\usepackage{setspace}
\usepackage[dvipsnames]{xcolor}
\usepackage{natbib}
\usepackage{indentfirst}
\usepackage{amsfonts}
\usepackage{graphicx}
\usepackage{subcaption}
\usepackage{natbib}
\usepackage{caption}
\usepackage{tablefootnote} 
\usepackage{makecell}
\usepackage{amssymb}
\usepackage{latexsym}
\usepackage{amsmath, amsfonts,amssymb, amsthm, euscript,makeidx,color,mathrsfs}
\usepackage{enumerate}
\usepackage[bookmarksopen=true]{hyperref}
\usepackage{bookmark}
\usepackage{booktabs}
\usepackage{listings}
\usepackage{appendix}
\usepackage{array}
\newcommand{\PreserveBackslash}[1]{\let\temp=\\#1\let\\=\temp} 
\newcolumntype{C}[1]{>{\PreserveBackslash\centering}p{#1}} 
\newcolumntype{R}[1]{>{\PreserveBackslash\raggedleft}p{#1}} 
\newcolumntype{L}[1]{>{\PreserveBackslash\raggedright}p{#1}} 
\hypersetup{colorlinks=true,linkcolor=blue}
\usepackage{listings}
\definecolor{dgray}{gray}{0.35} % colour of R comments
\definecolor{lgray}{gray}{0.95} % background colour of R-code

\hypersetup{colorlinks=true,linkcolor=blue,citecolor=gray}

\title{Scale-Location-Truncated Beta Regression: Expanding Beta Regression to Accommodate 0 and 1}
\author{Mingang Kim, Brent A. Kaplan, Mikhail N. Koffarnus, Christopher T. Franck}
\date{}

\begin{document}
\doublespacing
\maketitle
\section*{Abstract}

Beta regression is frequently used when the outcome variable y is bounded within a specific interval, transformed to the (0, 1) domain if necessary. However, standard beta regression cannot handle data observed at the boundary values of 0 or 1, as the likelihood function takes on values of either 0 or $\infty$. To address this issue, we propose the Scale-Location-Truncated beta (SLTB) regression model, which extends the beta distribution’s domain to the [0, 1] interval. By using scale-location transformation and truncation, SLTB distribution allows positive finite mass to the boundary values, offering a flexible approach for handling values at 0 and 1. In this paper, we demonstrate the effectiveness of the SLTB regression model in comparison to standard beta regression models and other approaches like the Zero-One Inflated Beta (ZOIB) mixture model \citep{liu2015zoib} and XBX regression \citep{kosmidis2025extended}. Using empirical and simulated data, we compare the performance including predictive accuracy of the SLTB regression model with other methods, particularly in cases with observed boundary data values for y. The SLTB model is shown to offer great flexibility, supporting both linear and nonlinear relationships. Additionally, we implement the SLTB model within maximum likelihood and Bayesian frameworks, employing both hierarchical and non-hierarchical models. These comprehensive implementations demonstrate the broad applicability of SLTB model for modeling data with bounded values in a variety of contexts.

\section{Introduction}
The beta regression model proposed by \cite{cribari2010beta} is widely used in situations where the outcome is bounded within an interval (a, b) because a simple linear transformation can map these values to (0, 1). Due to its wide use, we call this the standard beta regression model. However, standard beta regression  has a limitation: while it is possible to observe 0 and 1 values in data, standard beta regression cannot handle exact 0 and 1 values in the outcome variable. This is because the likelihood-based approach becomes problematic at these boundaries. The beta density at y=0 or y=1 is either 0 or $\infty$, except in certain cases such as when $\alpha=1, \ \beta=1 $. Consequently, the likelihood function evaluates to 0 or $\infty$, making likelihood-based methods usable for model fitting. This issue extends to standard beta regression. Specifically, the log likelihood function is undefined if any of the observed outcome data are 0 or 1, limiting the model's applicability when boundary observations are present. To address this issue, we propose the Scale-Location-Truncated beta (SLTB) regression model.

\begin{figure}[H]
    \centering
    \begin{subfigure}[b]{0.48\textwidth}
        \includegraphics[width=\textwidth]{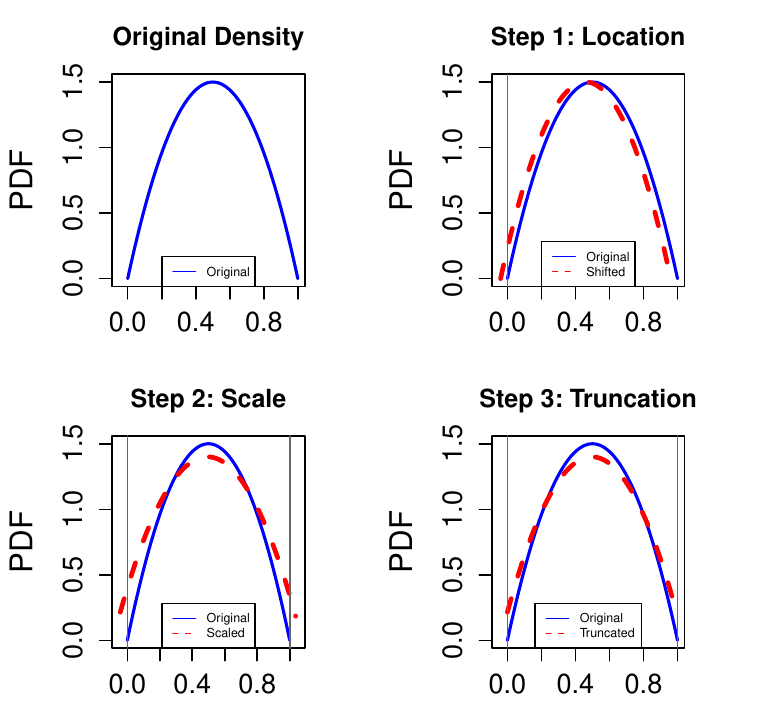}
        \caption{SLTB distribution with $\alpha$=2 and $\beta$=2 (*)}
    \end{subfigure}
    \hfill
    \begin{subfigure}[b]{0.48\textwidth}
        \includegraphics[width=\textwidth]{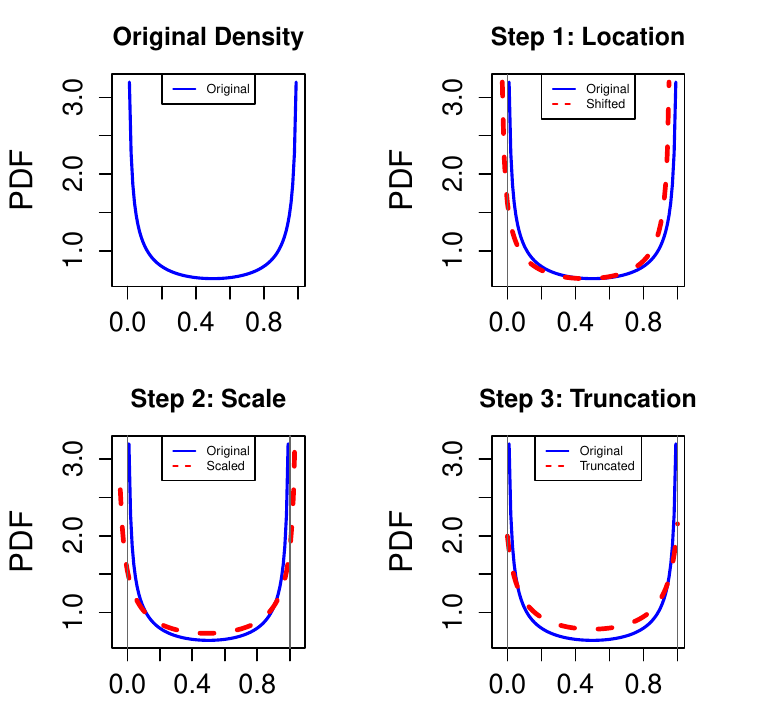}
        \caption{SLTB distribution with $\alpha$=0.5 and $\beta$=0.5 (*)}
    \end{subfigure}
    \caption{The top left panel shows the original beta probability density function. Step 1: Location, the beta density is shifted down the x-axis using location value $l$. Step 2: Scale, the adjusted density from Step 1 is scaled using scale value $s$. Step 3: Truncation, the density is truncated to have positive finite mass on [0, 1].  We intentionally selected sufficiently large values for the scale parameter ($s = 1.08$) and the location parameter ($l = 0.04$) to ensure that differences in the plot are visually distinguishable. (*) The parameters $\alpha$ and $\beta$ denote shape parameters.
 }
    \label{fig:slt_process}
\end{figure}

Figure~\ref{fig:slt_process} illustrates the basic concept behind the SLTB density. The panels show the step-by-step process translating the standard beta density to the SLTB density. First we extend the beta distribution by expanding its domain beyond [0,1] through a scale–location transformation, defined by standard scale value $s$ and location value $l$. Since this extended domain includes values outside the \([0, 1]\) interval, we next apply truncation to restrict the support back to \([0, 1]\). This truncation allows the SLTB density retains positive and finite probability at the boundary values of 0 and 1. 

Truncated distributions have played an important role in statistical modeling, especially in contexts where data is censored or naturally restricted. Early work by \cite{pearson1908generalised} discussed truncated normal distributions and explored how truncation affects estimates such as correlation coefficients. Later, \cite{johnson1995continuous} extended the theoretical framework by covering a broad class of truncated distributions, including truncated gamma and log-normal distributions. 

\begin{figure}[H]
  \centering
  \begin{subfigure}[b]{0.45\textwidth}
    \centering
    \includegraphics[width=\textwidth]{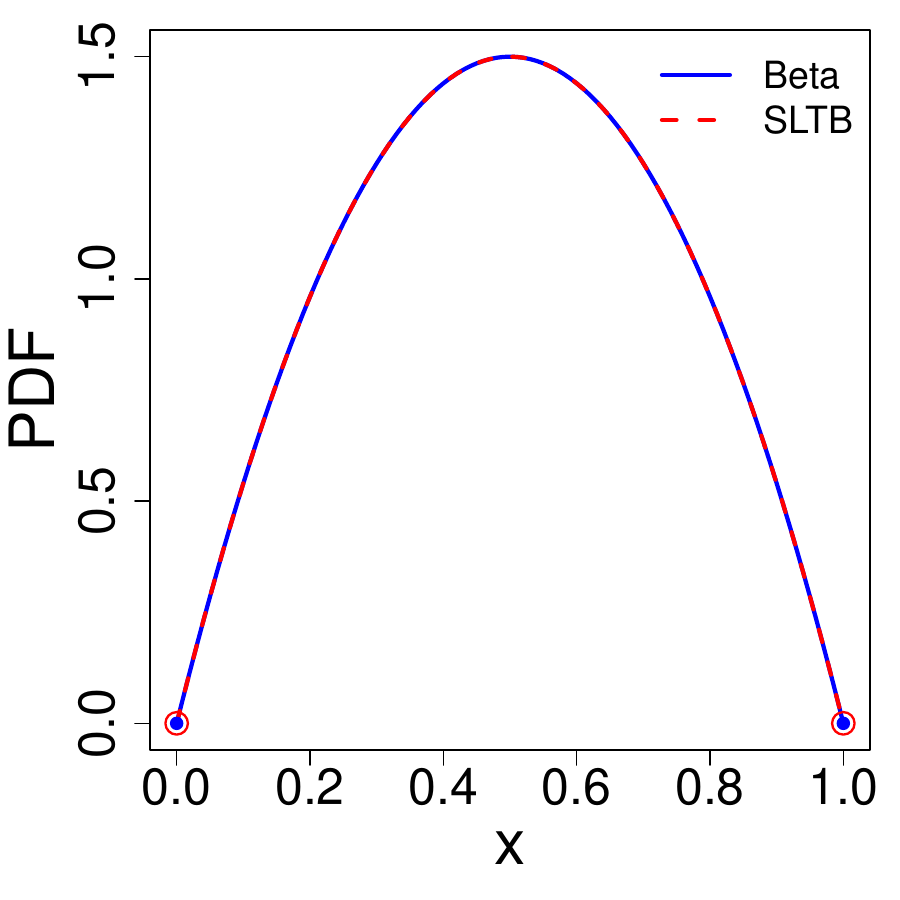}
  \end{subfigure}
  \hfill
  \begin{subfigure}[b]{0.45\textwidth}
    \centering
    \includegraphics[width=\textwidth]{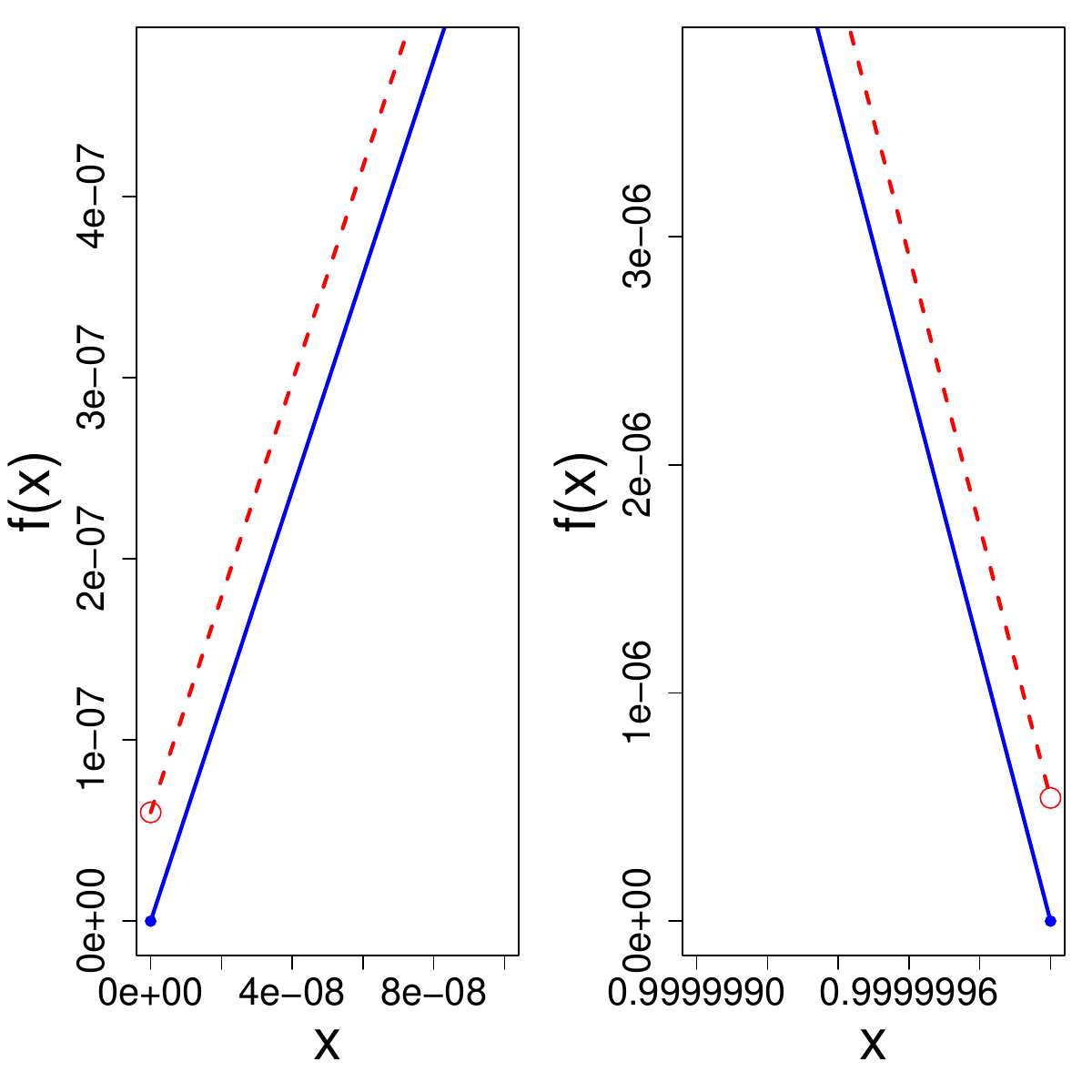}
  \end{subfigure}
  \caption{Left panel is the probability density function (PDF) of the SLTB and the beta distribution ($\alpha$=2, $\beta$=2) when $s = 1 + 10^{-8.5}$ and $l = 10^{-9}$—the same values used in the case studies presented in this paper. These density functions are visually indistinguishable. The right panel shows the region near 0 and 1. While the PDF of the beta distribution is 0 in when x is 0 and 1, the SLTB distribution has positive finite mass at the boundary.} 
  \label{fig:slt_density}
\end{figure}

We exploit an under-recognized and seldom-utilized aspect of truncated distributions, which is that they have positive and finite density on their boundary values. Thus the SLTB density is positive and finite on domain \([0, 1]\). This simple observation is the core technical insight that makes SLTB methods work. As shown in Figure \ref{fig:slt_density}, the resulting distribution closely approximates the beta distribution. By selecting appropriate scale $s$ and location $l$ values, our approach retains indistinguishably close fit to the beta distribution, offering an enhanced modeling framework for bounded data in the presence of boundary values. In addition, this allows the highly similar inference for data on (0, 1) between the standard beta regression and SLTB regression. The interpretation of parameters is the same between standard beta regression and SLTB regression, which is not the case for all competing methods.

Several alternative approaches have been developed to address the issue of observed 0 and 1 data values in context of beta regression. One such approach is the Zero-One Inflated Beta (ZOIB) model \citep{liu2015zoib}, which explicitly models point masses at 0 and 1. This mixture model separates the data into three components: one for values at 0, a second for values at 1, and a third for values strictly between 0 and 1. 
The XBX regression model \citep{kosmidis2025extended} was recently introduced as a linear modeling framework that accommodates boundary values at 0 and 1 by employing a continuous mixture of beta distributions. The XBX distribution extends the support to the closed [0,1] interval. An implementation of the XBX regression model is provided in the \texttt{betareg} package for \textsf{R} \citep{cribari2010beta}.  While this extension increases model flexibility by including boundary values, it is currently described in terms of and implemented only for linear, non-hierarchical models. In contrast, we show the SLTB regression model supports both linear and nonlinear relationships, providing a versatile solution that can be applied to a broader range of applications only requiring a relatively simple tweak to the standard beta distribution. This additional capability makes the SLTB model particularly useful when the data structure requires nonlinear modeling as in the behavior economic case study in Section \ref{sec: Bayesian nonlinear}.

In this paper, we introduce the SLTB model and demonstrate how it handles boundary values while preserving the flexibility of the beta distribution. By simply replacing the standard beta density with the SLTB density, we enable seamless inference in the presence of 0s and 1s across a wide range of beta models, including linear regression using maximum likelihood estimation (MLE) (Section \ref{sec: linear}), Bayesian hierarchical linear models (Section \ref{sec: Bayesian linear}) and Bayesian hierarchical nonlinear models (Section \ref{sec: Bayesian nonlinear}). 

We compare the SLTB model with existing approaches, including the Bayesian mixture-based ZOIB model \citep{liu2015zoib} and the continuous mixture-based XBX model \citep{kosmidis2025extended}, in datasets that contain boundary values at 0 or 1. Additionally, we demonstrate the SLT model's implementation in both classical and Bayesian frameworks, offering a comprehensive overview of its broad applicability in real-world data scenarios. The SLTB regression framework offers a parsimonious alternative to these mixture-based approaches, as it requires no more parameters than standard beta regression. By contrast, models such as the ZOIB include separate parameters for modeling boundary values at 0 and 1, as well as outcomes between 0 and 1. While this structure provides flexibility, it also increases model complexity. Although the SLTB regression model is relatively simple compared to these other methods, it performs favorably, showing comparable performance with faster computation speed in the data we analyze.

The rest of the paper is organized as follows. Section~\ref{sec: Methods} introduces the proposed methodology, including SLTB regression. Section~\ref{sec: Methodology application} applies the SLTB regression to both simulated and empirical data, comparing its performance with existing models discussed earlier. Section~\ref{sec: Discussion} concludes with a summary, discussion of limitations, and suggestions for future research.

\section{Methods}
\label{sec: Methods}

\subsection{Beta Distribution, Reparameterization, and a Boundary Problem}
The beta density for random variable y is expressed as Equation (\ref{Eq1})

\begin{equation}
    f(y; \alpha, \beta) = \frac{\Gamma{(\alpha+\beta)}}{\Gamma{(\alpha)}\Gamma{(\beta)}} y^{\alpha-1} (1-y)^{(\beta-1)}, \ \  0 < y < 1
    \label{Eq1}
\end{equation}

where $\alpha$ and $\beta$ are the shape parameters ($\alpha, \beta$ $>$ 0) and $\Gamma(\cdot)$ is the gamma function. The reparameterization suggested by \cite{ferrari2004beta} allows the beta distribution to be expressed in terms of its mean ($\mu$) and precision ($\phi$), where

\begin{equation}
    \begin{aligned}
        \mu &= \frac{\alpha}{\alpha+\beta}   \quad & 
\phi = \alpha+\beta. \quad 
    \end{aligned}
    \label{beta_density}
\end{equation}

\noindent This reparameterization allows the mean $\mu$ to be modeled as a function of predictors in regression problems, directly associating it with the predictors. In addition, the precision $\phi$ can be modeled as a function of covariates to quantify and characterize variability. The reparameterized beta density function for  y $\sim$ $B(\mu, \phi)$ is 

\begin{equation}
     f(y;\mu, \phi)= \frac{\Gamma(\phi)}{\Gamma(\mu\phi)\Gamma((1-\mu)\phi)}y^{\mu\phi-1}(1-y)^{(1-\mu)\phi-1}, \quad 0 < y < 1,
     \label{Eq2}
\end{equation}

\noindent where $\text{E}(y)=\mu$ and $\text{Var}(y)= \frac{\mu(1-\mu)}{1+\phi}$, with $0 < \mu < 1$ and $\phi > 0$.

Based on the reparameterization (\ref{beta_density}), the likelihood function for independent data is 

\begin{equation*}
   L(y_{i};\mu, \phi)=\prod_{i}^{N} \frac{\Gamma(\phi)}{\Gamma(\mu\phi) \Gamma((1-\mu)\phi)} y_{i}^{\mu\phi-1}(1-y_{i})^{(1-\mu)\phi-1}.
\end{equation*}

Since the beta density at y=0 and y=1 is either 0 or $\infty$ in most cases, the likelihood function (\ref{likelihood}) for the data goes to 0 or $\infty$, making parameter $\mu$ and $\phi$ impossible to estimate using likelihood-based methods. Even a single 0 or 1 values for y makes the model become unusable without modifications, necessitating approaches to handle boundary observations effectively.

The log-likelihood function for the data is 
\begin{equation}
\begin{aligned}
   l(y_{i};\mu, \phi) &= ln(L(y_{i};\mu, \phi)) \\
   &= \sum_{i}^{N} \Big[ ln(\Gamma(\phi)) - ln(\Gamma(\mu\phi)) - ln(\Gamma((1-\mu)\phi)) \\
   &\quad + (\mu\phi - 1)ln(y_{i}) + ((1-\mu)\phi - 1)ln(1 - y_{i}) \Big].
\end{aligned}
   \label{likelihood}
\end{equation}

In the same context, the log-likelihood function cannot be used to maximize the likelihood when the data has 0 or 1 value because $ln(0)$ and $ln(\infty)$ are undefined. Thus, likelihood-based approach is not possible directly from standard beta regression when there exists 0 or 1 value in dataset. 

\subsection{Standard Beta Regression}

Standard beta regression as described by \cite{cribari2010beta} is based on the reparameterized beta distribution (\ref{beta_density}), and it can accommodate linear and nonlinear models for continuous $y$ in the open interval (0, 1).

Let random variable  $y_i \sim B(\mu_i, \phi_i)$ (i=1, 2, $\cdots$, N). By using the reparameterization, standard beta regression model can be defined as

\begin{equation*}
    g(\mu_i) = \eta_i = x_i^T\beta 
\end{equation*}
where $x_i=(x_{i1}, x_{i2}, \dots, x_{ik})^T$ is  $k \times 1$ a vector of regressor (k $<$ n) and $\beta = (\beta_1, \beta_2, \dots, \beta_k)^T$ is $k \times 1$ vector of regression parameters.
Here, $g(\cdot)$ : (0, 1) $\rightarrow \  \mathbb{R}$  is a link function. In the absence of 0s and 1s, inference for standard beta regression can proceed by maximum likelihood estimation. 

\subsection{Scale-Location-Truncated Beta (SLTB) Distribution}
\label{sec: SLTB_dist}
By using a scale value $s$ and a location value $l$, the beta density can be modified into the SLTB density which accommodates exact 0 and 1 values. The values $s$ and $l$ are tuning parameters chosen to make the SLTB density closely resemble the beta density in the interval $(0,1)$ while placing finite mass on the endpoints as shown in right panel of Figure \ref{fig:slt_density}. In this paper, we selected the scale value $s=1+10^{-8.5}$ and location value $l=10^{-9}$. These values minimized squared differences on a fine grid between the SLTB density and standard beta density when $\alpha=\beta=2$.         

First, we apply a scale-location transformation using a scale value $s$ and location value $l$ as in the Step 1 and Step 2 from the Figure \ref{fig:slt_process}. 

Let $z$ be scale-location transformed form of $y$, where $y$ follows the beta distribution as in Equation (\ref{Eq2}). That is, the random variable $z$ = $(y-l)\cdot s$. The resulting scale-location transformed density for $z$ is as follows.  

\begin{equation}
    f_Z(z;\mu, \phi, s ,l)=\frac{1}{s} \frac{\Gamma(\phi)}
{\Gamma(\mu\phi)\Gamma((1-\mu)\phi)}\left(\frac{z}{s}+l\right)^{\mu\phi-1}\left(1-\left(\frac{z}{s}+l\right)\right)^{(1-\mu)\phi-1}
\label{Eq3}
\end{equation}

\noindent where $s$ is scale value and $l$ is location value. $\mu, \phi$ are shape parameters and $\Gamma$ is a gamma function. 

This scale-location transformation expands the domain of the density (\ref{Eq3}) to [$-ls, (1-l)s$], making $f_Z$ extend slightly beyond the original domain (0, 1). Thus, it has positive mass outside the targeted [0, 1] support.

Then as in Step 3 of Figure \ref{fig:slt_process}, we truncate scale-location transformed density $f_Z$ in Equation (\ref{Eq3}). 

We follow the descriptions of truncated distributions in \cite{nadarajah2006r}, 

\begin{equation}
    f_X(x) =
    \begin{cases} 
        \frac{g(x)}{G(b) - G(a)}, & \text{if } a \leq x \leq b, \\
        0, & \text{otherwise},
    \end{cases}
    \label{eq:truncation}
\end{equation}

\hspace{5pt}where $g(\cdot)$ is the probability density function (pdf) of a distribution, and $G(\cdot)$ is its cumulative distribution function (CDF).

 To restrict the domain of $z$ to [0, 1], we truncate the Equation (\ref{Eq3}) following the Equation (\ref{eq:truncation}).

The final form of the SLTB distribution is 

\begin{equation}
    f_G(g)= \frac{f_Z(g)}{F_z(1)-F_z(0)}=\frac{\frac{\Gamma(\phi)}
{\Gamma(\mu\phi)\Gamma((1-\mu)\phi)}\left(\frac{g}{s}+l\right)^{\mu\phi-1}\left(1-\left(\frac{g}{s}+l\right)\right)^{(1-\mu)\phi-1}\frac{1}{s}}{F_z(1)-F_z(0)}. \ \ 0\leq g \leq 1.
\label{eq: SLTB density}
\end{equation} 

\noindent where $F_{z}(\cdot)$ is CDF.

The mean and variance of the SLTB density (\ref{eq: SLTB density}) are

\begin{equation}
    \begin{aligned}
        E(G) &= s \left[\mu-l\right], \quad & VAR(G) &= s^2 \left[\frac{\mu(1-\mu)}{\phi+1}\right].
    \end{aligned}
    \label{Eq4}
\end{equation}

As $s$ goes to 1 and $l$ goes to 0, the mean and the variance in Equation (\ref{Eq4}) converge to the mean and variance of the beta density. This is formally shown in the Online Supplement. The comparison of the mean and variance between standard beta density and SLTB density can be found in the Online Supplement as well.

\subsection{Scale-Location-Truncated Beta (SLTB) Regression}

The SLTB regression model is based on the SLTB distribution (\ref{sec: SLTB_dist}), and it can accommodate linear and nonlinear models for continuous $y$ in the closed interval [0, 1].

Let the random variable $y_i \sim SLTB(\mu_i, \phi_i)$ for $i = 1, 2, \dots, N$. 
Using the reparameterization, the SLTB regression model can be defined as
\begin{equation*}
   g(\mu_i) = \eta_i = f(\mathbf{x}_i, \boldsymbol{\beta}),
\end{equation*}
where $\mathbf{x}_i = (x_{i1}, x_{i2}, \dots, x_{ik})^T$ is a $k \times 1$ vector of regressors ($k < N$), 
and $\boldsymbol{\beta} = (\beta_1, \beta_2, \dots, \beta_k)^T$ is a $k \times 1$ vector of regression parameters. 
Here, $g(\cdot) : (0,1) \rightarrow \mathbb{R}$ is a link function. The function $f(\mathbf{x}_i, \boldsymbol{\beta})$ may represent a linear, nonlinear, hierarchical linear, or hierarchical nonlinear model. 
Additional indexing for specific types of hierarchical models will be introduced later.

The log-likelihood function for the data is 

\begin{equation*}
\begin{aligned}
   l(y_{i};\mu, \phi) 
   &= \ln \big( L(y_{i};\mu, \phi) \big) \\
   &= \sum_{i=1}^{N} \Big[ 
        \ln \Gamma(\phi) 
        - \ln \Gamma(\mu\phi) 
        - \ln \Gamma\!\big((1-\mu)\phi\big) \\
   &\quad + (\mu\phi - 1)\,\ln\!\left(\frac{y_{i}}{s} + l \right) 
        + \big((1-\mu)\phi - 1\big)\,\ln\!\left(1 - \left(\frac{y_{i}}{s} + l \right)\right) \\
   &\quad - \ln\!\big(s\,(F_z(1) - F_z(0))\big) 
   \Big].
\end{aligned}
\end{equation*}

\section{Applications of SLTB Regression and Comparison with Existing Methods}
\label{sec: Methodology application}

In this section, we demonstrate the application of SLTB regression across various contexts: (i) linear beta regression using Maximum Likelihood Estimation (MLE), compared with the Bayesian mixture–based ZOIB model and the continuous mixture–based XBX model (Section~\ref{sec: comparison}); (ii) a Bayesian hierarchical linear model (Section~\ref{sec: Bayesian linear}); and (iii) a Bayesian hierarchical nonlinear model (Section~\ref{sec: Bayesian nonlinear}).

\subsection{Data Analytic Comparison of SLTB, ZOIB and XBX Regression Model}
\label{sec: comparison}

\subsubsection{\textit{ReadingSkills} data}

We analyzed reading accuracy and non-verbal IQ data from the \textit{ReadingSkills} dataset, available in the \texttt{betareg} R package \citep{cribari2010beta}. The dataset consists of 44 observations, with two versions of reading accuracy (accuracy and accuracy1), dyslexia diagnosis, and non-verbal IQ variables. Reading accuracy is proportional data bounded in [0, 1]. While the maximum possible value of the variable accuracy is restricted to 0.99, accuracy1's maximum possible value is 1. In other words, these are identical data except in cases where accuracy has a 0.99 value, accuracy1 has a 1 value. We analyze the data with boundary cases (accuracy1) in this paper. We compare SLTB with with standard beta regression for the non-boundary data (accuracy) in the Online Supplement.

 IQ is a numerical variable indicating intelligence quotient transformed to z-scores. Dyslexia diagnosis is represented as a binary factor indicating whether a child is diagnosed as dyslexic. We compare the SLTB model, XBX model estimated with the \texttt{betareg} R package, and the ZOIB model, estimated with the \texttt{ZOIB} R package \cite{liuRpack}, using \texttt{accuracy1} as the outcome variable. 

We adopt the following model:

\begin{equation}
    g(\mu_i) = \text{logit}(\mu_i) = \beta_0 + \beta_1 \cdot \text{Dyslexia}_i + \beta_2 \cdot \text{IQ}_i + \beta_3 \cdot (\text{Dyslexia}_i \times \text{IQ}_i).
    \label{eq:logit_model}
\end{equation}

For the accuracy outcome variable, \texttt{betareg} estimates parameters using standard beta regression via a likelihood-based approach (\ref{likelihood}). In contrast, when boundary values are present, i.e., using accuracy1 as an outcome, \texttt{betareg} employs XBX regression \citep{kosmidis2025extended} as an alternative to handle these values effectively.

\begin{figure}[H]
    \centering
    \includegraphics[width=0.8\textwidth]{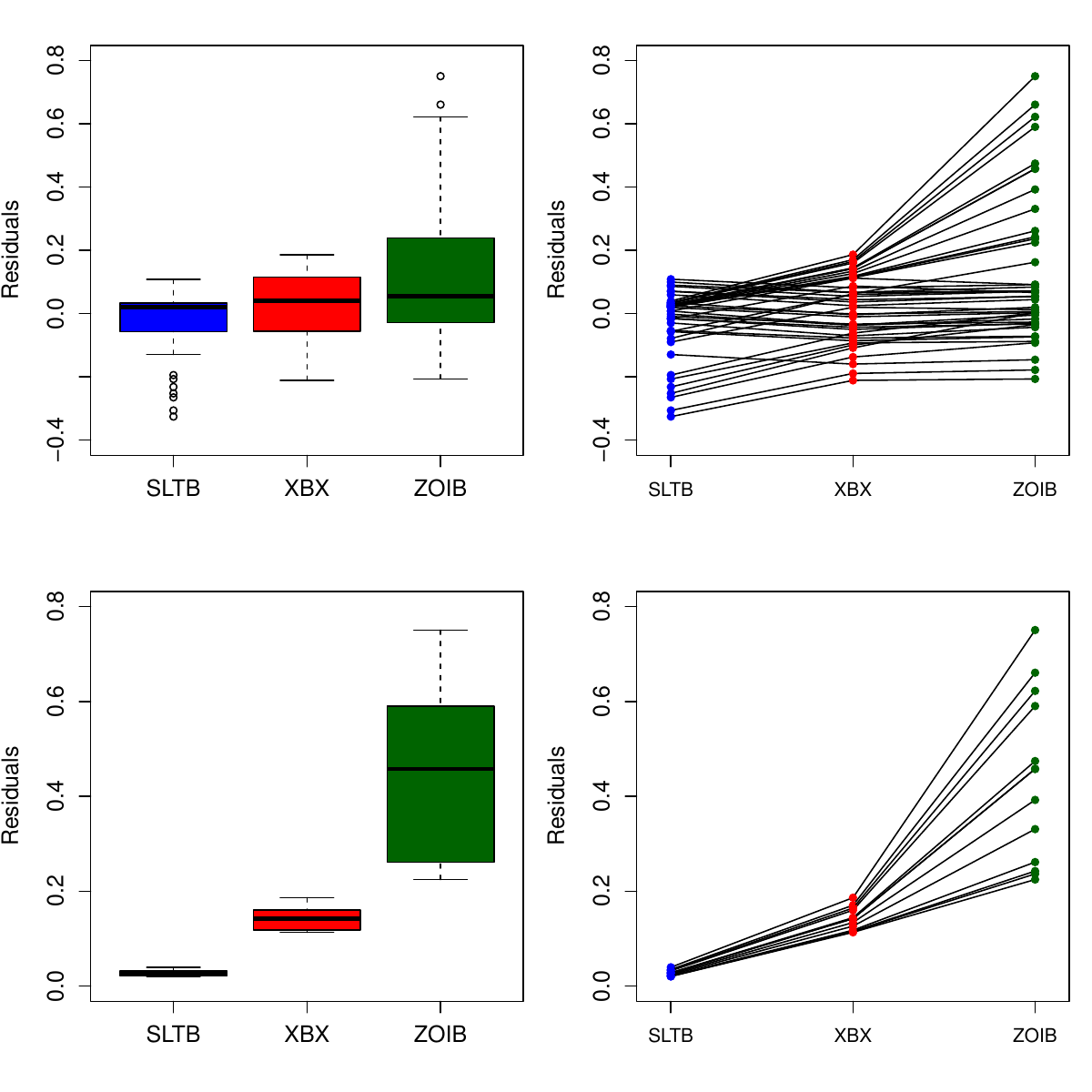}
    \caption{The top left panel shows boxplots of residuals across all subjects for the three models. The SLTB model (blue) yields residuals centered closer to zero with relatively small variability. The XBX regression (red) produces slightly larger residuals, while the ZOIB model (green) displays the largest spread and median, indicating a less precise fit overall. The top right panel connects individual residuals across models. Many of the lines slope upward from SLTB to XBX and further to ZOIB, suggesting that residuals consistently increase as we move from SLTB to XBX to ZOIB. The bottom left panel focuses on the subset of subjects with an outcome equal to 1. Here, the SLTB residuals are tightly clustered near zero, indicating an excellent fit for this group. XBX residuals are larger, and ZOIB residuals are markedly higher with much greater spread. The bottom right panel shows paired residuals for the same subset, again highlighting the pattern: SLTB consistently produces the smallest residuals, followed by XBX, with ZOIB yielding the largest residuals.}
    \label{fig:slt_xbx_zoib_combined}
\end{figure}

Based on the SLTB regression model, SLTB regression implementation used a maximum likelihood-based approach to fit to the model in Equation (\ref{eq:logit_model}). 
We used the mean squared error (MSE) to illustrate the relative performance of the three models. The SLTB regression model achieved an MSE of 0.0130, while the XBX model performed slightly better with an MSE of 0.0114. In contrast, the ZOIB model had a much larger MSE of 0.0696, reflecting poorer overall predictive accuracy. For the subset of boundary values (i.e., outcomes equal to 1), the differences between models were even more pronounced. The SLTB model achieved the smallest error of 0.0008, indicating an exceptionally good fit for these cases. The XBX model yielded a higher error of 0.0204, while the ZOIB model performed substantially worse with an error of 0.2196. Figure~\ref{fig:slt_xbx_zoib_combined} visualizes these results by displaying residual distributions across models. Consistent with the MSE values, SLTB residuals are centered closer to zero and exhibit the least variability. XBX residuals are somewhat larger, while ZOIB residuals are markedly more dispersed and shifted upward. For the boundary value cases in particular, SLTB residuals cluster tightly around zero, XBX residuals spread more widely, and ZOIB residuals display the largest deviations. The ZOIB model's poorer performance on these data is due to a separation issue analogous to those in binary regression problems. Boundary values of 1, corresponding to a perfect reading score, are only observed in the non-dyslexia group. This seems to causes the boundary component of the ZOIB model to have unstable coefficients and poorer performance. See Section \ref{sec: Discussion} for further discussion.

Table \ref{tab:xbx_slt_zoib} shows estimates, standard errors, and test results for the SLTB, XBX, and ZOIB models. We note that the interpretation of parameters among the three models is different, thus we do not expect them to necessarily have similar estimates. We also note that the standard errors for SLTB are higher than XBX. However, the SLTB standard errors are closer than XBX to results from standard beta regression in the absence of boundary values for this data set, see Online Supplement.

\begin{table}[H]
\centering
\begin{tabular}{lrrrr}
  \hline
  \multicolumn{5}{c}{\textbf{SLTB}} \\
  \hline
    \textbf{Variable} & \textbf{Estimate} & \textbf{Std. Error} & \textbf{z value} & \textbf{p-value} \\ 
  \hline
  (Intercept) & 1.8622 & 0.1996 & 9.327 & $<$ 0.001  \\ 
  dyslexia    & -1.5442 & 0.1889 & -8.176 & $<$ 0.001  \\ 
  iq          & 0.0978 & 0.1493 & 0.655 & 0.5126 \\ 
  dyslexia:iq & -0.1487 & 0.1494 & -0.996 & 0.3194 \\ 
  \hline
  \multicolumn{5}{c}{\textbf{XBX}} \\
  \hline
  (Intercept) & 0.3228 & 0.0357 & 9.043 & $<$ 0.001  \\ 
  dyslexia    & -0.1895 & 0.0241 & -7.874 & $<$ 0.001  \\ 
  iq          & 0.0346 & 0.0199 & 1.738 & 0.0822 \\ 
  dyslexia:iq & -0.0411 & 0.0200 & -2.050 & 0.0404 \\ 
    \hline
  \multicolumn{5}{c}{\textbf{ZOIB}} \\
 \hline
   \textbf{Variable} & \textbf{Posterior mean} & \textbf{Posterior SD} & & \\ \hline
   \multicolumn{5}{l}{\textit{Coefficient for (0, 1) value}}\\
   (Intercept) & 0.8817  & 0.1065 & - & -  \\ 
  dyslexia    & -0.5030 & 0.1074 & - & -  \\ 
  iq          & 0.2879 & 0.1276 & - & -  \\ 
  dyslexia:iq & -0.3480  & 0.1275 & - & -  \\ 
  \hline
    \multicolumn{5}{l}{\textit{Coefficient for boundary value}}\\
   (Intercept) & -15.8209 & 10.3072 &  - & -   \\ 
  dyslexia    & -15.4802 & 10.3062 &  - & -  \\ 
  iq          & 1.5113  & 7.9662 &  - & -  \\ 
  dyslexia:iq & 0.6570  & 7.9668  &  - & -  \\ 
  \hline
\end{tabular}
\caption{Comparison of Coefficient Estimates from SLTB, XBX, and ZOIB Regression Models.
This table reports parameter estimates, standard errors, z-values, and p-values for the SLTB and XBX models, each using a logit link for the mean structure. For the Bayesian ZOIB model, uncertainty is expressed through the posterior standard deviation, which serves as the analogue to the frequentist standard error. The results highlight notable differences in both the magnitude and statistical significance of coefficients across models.}
\label{tab:xbx_slt_zoib}
\end{table}

\subsubsection{Simulation data}

We used a simulation study to evaluate the model fit and computational efficiency of SLTB regression in comparison with two alternative approaches: Zero-One Inflated Beta (ZOIB) regression \citep{liu2015zoib} and XBX regression \citep{kosmidis2025extended} for boundary values. Simulation results revealed clear differences between the methods, highlighting the computational efficiency and accuracy of the SLTB regression model despite its relative simplicity. Both ZOIB and XBX regression models are implemented with linear predictor structures in their respective papers and R packages. To ensure comparability, we also restricted our analysis to linear predictors. Notably, the \texttt{betareg} package, when boundary values are present, relies on XBX regression \citep{kosmidis2025extended} to handle such cases effectively.

The analysis examined model performance under varying sample sizes, with particular attention to scenarios involving boundary observations ($Y = 1$). A total of 1,000 Monte Carlo replications were performed for each setting, with datasets generated at either $N = 20$ (small sample) or $N = 400$ (large sample) observations. Data were simulated following the structure of the \textit{ReadingSkills} dataset from the \texttt{betareg} R package \citep{cribari2010beta}, which includes a binary indicator of Dyslexia and a continuous measure of IQ. Specifically, the binary variable, $X_1$, corresponding to Dyslexia, was recoded from its original \textit{yes}/\textit{no} factor levels to $0$ and $1$, with equal probability. The continuous variable, $X_2$, corresponding to IQ, was drawn from a normal distribution.

\[
N(\mu = 100, \ \sigma = 15)
\] 

and then standardized. 

The response variable was simulated using a beta regression framework, with the mean modeled via a logistic link function. The linear predictor included an intercept, the main effects of group and IQ, and their interaction:  
\[
    \eta_i = \beta_0 + \beta_1 \cdot X_{i1} + \beta_2 \cdot X_{i2} + \beta_3 \cdot X_{i1}X_{i2}, \; \; i = 1,\cdots, N
\]

where the true coefficients were set to:  
\[
    \beta_0 = 1.2, \quad \beta_1 = -0.88, \quad \beta_2 = 0.43, \quad \beta_3 = -0.52.
\]  
The mean response was then obtained via the inverse logit transformation:  
\[
    \mu_i = \frac{\exp(\eta_i)}{1 + \exp(\eta_i)}.
\]  

To induce a higher proportion of boundary values compared to the original \textit{ReadingSkills} dataset, the true coefficient values were deliberately chosen to produce boundary outcomes, and the precision parameter was fixed at $\phi = 10$. The response variable was then generated from a beta distribution $\alpha=\mu_i \cdot \phi$ and $\beta=(1 - \mu_i) \cdot \phi$,
ensuring $Y \in [0,1]$.  

In the small-sample scenario with boundary values included ($N = 20$), 731 Monte Carlo replications were available, since only those datasets contained boundary values under the data-generating process. In this setting, the mean squared error (MSE) of the SLTB regression model was 0.0136, while that of the XBX regression model was 0.0130 (Table~\ref{tab:combined_results}). In contrast, the ZOIB model exhibited a substantially higher MSE of 0.0357, indicating inferior model performance quantified by MSE when boundary values exist in the response variable and the sample size is small. In terms of computational efficiency, the SLT model demonstrated the lowest average runtime at 0.0110 seconds, followed by XBX (0.3375 seconds), while ZOIB was markedly slower at 17.0755 seconds.

When the sample size increased to 400 ($N = 400$, 1000 replications) under boundary conditions, the XBX model achieved the lowest mean squared error (MSE) at 0.0159, only slightly lower than SLT (0.0161). By contrast, the ZOIB model exhibited substantially higher error (0.0415). With respect to computation, SLTB remained the fastest method at 0.0540 seconds on average, while XBX and ZOIB required 0.8431 and 229.19 seconds, respectively. 

Taken together, these results highlight the effectiveness and efficiency of the SLTB regression model across a range of data-generating scenarios. SLTB consistently maintained low prediction error, particularly in small-sample and boundary-inclusive settings, while achieving substantial gains in computational speed relative to both XBX and ZOIB. Unlike ZOIB, which deteriorated in both accuracy and runtime under boundary conditions, SLTB demonstrated stable performance and scalability. Although XBX achieved slightly lower MSEs in certain cases, it showed higher computational cost potentially making it less practical in time-sensitive or resource-constrained applications. Therefore, the SLTB regression model represents a computationally efficient and statistically reliable alternative for modeling bounded outcomes.

\begin{table}[H]
\centering
\begin{tabular}{llcc}
\hline
 \textbf{$N$} & \textbf{Method} & \textbf{MSE} & \textbf{Time (s)} \\
\hline
 20   & SLTB  & 0.0136 & 0.0110 \\
      & XBX  & 0.0130 & 0.3375 \\
      & ZOIB & 0.0357 & 17.076 \\
\hline
400  & SLTB  & 0.0161 & 0.0540  \\
     & XBX  & 0.0159 & 0.8431  \\
     & ZOIB & 0.0415 & 229.19  \\
\hline
\end{tabular}
\caption{MSE and average computation time across different sample sizes when at least one response is equal to 1.}
\label{tab:combined_results}
\end{table}

The relatively poor performance of the ZOIB model, compared with XBX and SLTB, can be attributed to the role of $X_1$ as a predictor. In our simulation study, $X_1$ is inspired by the \textit{Dyslexia} variable from the \textit{ReadingSkills} dataset. The true coefficient in the simulation is chosen such that we only observed perfect accuracy scores of 1 in the dyslexia\textit{= no} condition. Similarly, the \textit{ReadingSkills} dataset includes perfect accuracy scores of 1 only in the dyslexia\textit{= no} group. This is analogous to a separation problem for binary regression. As a result, when ZOIB attempts to model the boundary component, it has no information for the Dyslexia \textit{= yes} group. This imbalance caused the coefficients of the submodel for boundary value to be unstable, thereby reducing ZOIB’s predictive performance relative to XBX and SLTB. However, this does not indicate a flaw in our data-generation process but rather highlights a structural limitation of ZOIB under certain conditions. Specifically, boundary on the relatively poor performance of the ZOIB model, compared with XBX and SLTB, can be attributed to the role of $X_1$ as a predictor. Analysts must be cautious; this analysis shows separation problems are plausible in real world applications, which complicate analysis in a mixture approach such as ZOIB. By contrast, SLTB and XBX do not suffer in these cases.

\subsection{A Comparative Study of ZOIB and SLTB Regression within Bayesian Hierarchical Linear Model Framework}
\label{sec: Bayesian linear}

Following the description and notation of \cite{liu2015zoib}, we analyzed student-level monthly alcohol use data. The description about this dataset can be found from \textit{AlcoholUse} dataset in the \texttt{ZOIB} R package \citep{liuRpack}. The dataset includes proportion of students who consumed alcohol (in the past 30 days) as its outcome variable. These proportions are provided by grade, gender, county,  and day range (0, 1–2, 3–9, 10–19, 20–30).  In our Bayesian hierarchical model, we use grade, gender, county, and midpoint of the day range as predictors. We incorporate county-level variation as a random effect to account for differences across counties. Since this example involves hierarhcical models we compare with the ZOIB approach but not XBX, since XBX is not currently available for hierarchical models.

Using the \texttt{ZOIB} R package, we implemented a Bayesian hierarchical linear model following the specific equation from \citep{liu2015zoib} :

\begin{equation}
\begin{aligned}
\text{logit}\left(\mu_{ij}\right) &= (\beta_{1,0} + u_i) + \beta_{1,1} \ x_{ij1} + \beta_{1,2} \ x_{ij2} + \beta_{1,3} \ x_{ij3} + \beta_{1,4} \ x_{ij4} + \beta_{1,5} \ x_{ij5} + \beta_{1,6} \ x_{ij6} \\
\log(\phi_i) &= \eta \\
\text{logit}(p_{ij}) &= \beta_{2,0} + \beta_{2,1} \ x_{ij1} + \beta_{2,2} \ x_{ij2} + \beta_{2,3} \ x_{ij3} + \beta_{2,4} \ x_{ij4} + \beta_{2,5} \ x_{ij5} + \beta_{2,6} \ x_{ij6} \ x_{ij6}
\end{aligned}
\label{zoib_eq}
\end{equation}

where \(u_i\) is a county-level random effect (with \(i = 1, \dots, 56\) counties), \(j\) denotes the \(j\)-th individual within each county and $x_{ijk}$ is \(k\)-th predictor variable for \(j\)-th individual in \(i\)-th county.

The parameter \({\beta}_{1,i}\)  represents the regression coefficient vector for gender, grade, and the day ranges and interaction between grade and gender.
\({\beta}_{2,i}\) captures these variables for a second set of model parameter. The random effect \(u_i\) is assumed to follow a normal distribution with mean 0 and precision \(\sigma^{-2}\). 

The prior specifications follow the setting described in \citep{liu2015zoib}, with the following distributions:

\begin{equation}
\begin{aligned}
\beta_{1,0} &\sim \mathcal{N}(0, 10^{3}), \quad &\beta_{2,0} &\sim \mathcal{N}(0, 10^{3}) \\
\beta_{1,k} &\sim \mathcal{N}(0, 10^{3}), \quad &\beta_{2,k} &\sim \mathcal{N}(0, 10^{3}) \quad \text{for } k = 1, \dots, 6 \\
\eta &\sim \mathcal{N}(0, 10^{3}), \quad &\sigma &\sim \text{Uniform}(0, 20)
\end{aligned}
\label{zoib_prior}
\end{equation}

where k is the number of predictors. 

We compared the ZOIB \citep{liu2015zoib} following the Equation (\ref{zoib_eq}) using the \texttt{ZOIB} package in R, with our SLTB regression framework and estimated via the Metropolis-Hastings algorithm with the same prior structure as Equation (\ref{zoib_prior}). 

The model used for SLTB regression framework is as follows:
\begin{equation*}
\begin{aligned}
\text{logit}\left(\mu_{ij}\right) &= (\beta_{0} + u_i) + \beta_{1} \ x_{ij1} + \beta_{2} \ x_{ij2} + \beta_{3} \ x_{ij3} + \beta_{4} \ x_{ij4} + \beta_{5} \ x_{ij5} + \beta_{6} \ x_{ij6}\\
\log(\phi_i) &= \eta \\
\end{aligned}
\end{equation*}

SLTB regression does not include a $\text{logit}(p_{ij})$ term because it employs a single model to fit the data, whereas ZOIB uses a mixture model. A potential advantage of the SLTB regression model is that the parameter interpretations are similar to standard beta regression and the number of parameters included is smaller, while the ZOIB approach has mixture component-type interpretations for parameters depending on whether points are on the boundary, and the number of parameters is larger.

While the SLTB regression model provides a single estimated parameter value for each predictor variable, the ZOIB model yields two estimated parameter values: one for the (0, 1) domain of the response variable, and another for when the response variable equals 0. The estimated parameters can be found in Table \ref{tab:BH_zoib}.

\begin{table}[H]
\centering
\begin{tabular}{llllll}
\hline

\multicolumn{6}{c}{\textbf{SLTB}}\\ 
\hline
\textbf{Effect} & \textbf{Parameter} & \textbf{Mean} & \textbf{Q1} & \textbf{Median} & \textbf{Q3} \\ \hline
Intercept & $\beta_{0}$ & -3.617 & -3.691 & -3.617 & -3.542 \\
Gender M & $\beta_{2}$ & -0.269 & -0.315 & -0.269 & -0.217 \\
Grade 9 & $\beta_{3}$ & 0.753 & 0.709 & 0.753 & 0.794 \\
Grade 11 & $\beta_{4}$ & 1.024 & 0.979 & 1.024 & 1.066 \\
MedDays & $\beta_{1}$ & -0.770 & -0.787 & -0.770 & -0.753 \\
Grade 9*Gender M & $\beta_{5}$ & 0.068 & 0.006 & 0.068 & 0.129 \\
Grade 11*Gender M & $\beta_{6}$ & 0.266 & 0.203 & 0.266 & 0.322 \\
\hline
$\eta$ &  & 3.603 & 3.572 & 3.603 & 3.631 \\
$\sigma^2$ &  & 0.535 & 0.462 & 0.535 & 0.623 \\
\hline
\multicolumn{6}{c}{\textbf{ZOIB}}\\
\hline

Intercept & $\beta_{1,0}$ & -3.467 & -3.551 & -3.468 & -3.386 \\
Gender M & $\beta_{1,1}$ & -0.052 & -0.148 & -0.050 & 0.041 \\
Grade 9 & $\beta_{1,2}$ & 0.703 & 0.608 & 0.703 & 0.788 \\
Grade 11 & $\beta_{1,3}$  & 0.955 & 0.870 & 0.957 & 1.039 \\
MedDays & $\beta_{1,4}$  & -0.822 & -0.860 & -0.822 & -0.786 \\
Grade 9*Gender M & $\beta_{1,5}$ & -0.127 & -0.247 & -0.126 & -0.000 \\
Grade 11*Gender M & $\beta_{1,6}$  & 0.048 & -0.073 & 0.043 & 0.170 \\
\hline
Intercept & $\beta_{2,0}$  & -3.032 & -3.666 & -3.039 & -2.512 \\
Gender M & $\beta_{2,1}$ & 0.487 & -0.257 & 0.465 & 1.218 \\
Grade 9 & $\beta_{2,2}$ & -0.511 & -1.489 & -0.458 & 0.335 \\
Gender 11 & $\beta_{2,3}$ & -0.906 & -2.117 & -0.869 & 0.177 \\
MedDays & $\beta_{2,4}$ & 0.255 & 0.009 & 0.257 & 0.548 \\
Grade 9*Gender M & $\beta_{2,5}$ & -0.337 & -1.711 & -0.339 & 0.927 \\
Grade 11*Gender M & $\beta_{2,6}$ & -0.690 & -2.366 & -0.656 & 0.992 \\
\hline
$\eta$ & & 4.385 & 4.302 & 4.386 & 4.461 \\
$\sigma^2$ & & 0.021 & 0.012 & 0.020 & 0.034 \\
\hline
\end{tabular}

\caption{Posterior means and quantiles (Q1, median, and Q3) for two models: the Zero-One Inflated Beta (ZOIB) model and the Bayesian hierarchical linear model using the SLTB distribution. Both models capture the effects of grade, gender, and MedDays on the intercepts and slopes for two groups. In the ZOIB model, parameters $\beta_{1,i}$ represent the estimates for the (0, 1) portion, and parameters $\beta_{2,i}$ represent the estimates for the zero-inflation portion. In contrast, the SLTB model provides a single estimate for each variable.}

\end{table}
\label{tab:BH_zoib}

\begin{figure}[H]
    \centering
    \includegraphics[width=0.8\linewidth]{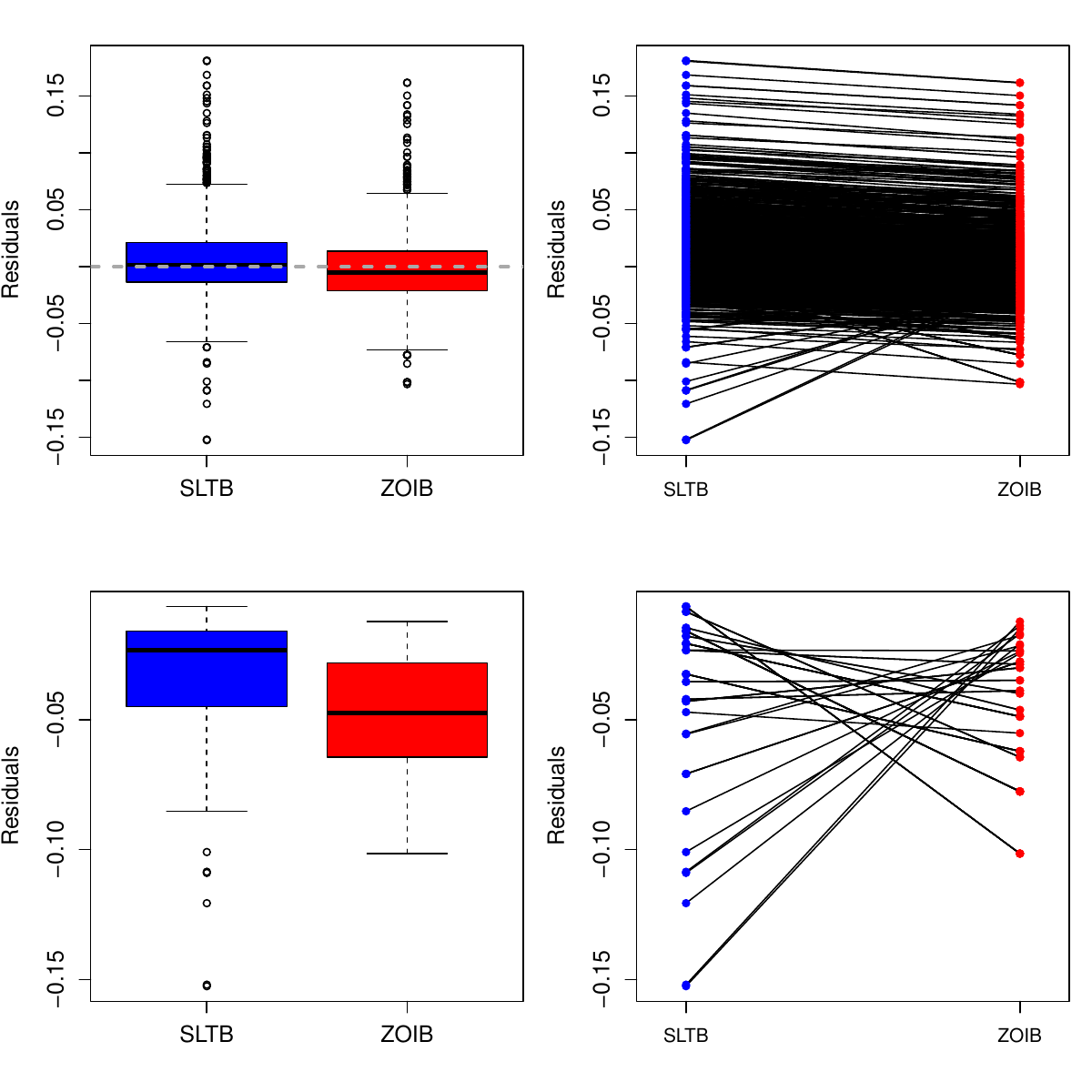}
     \caption{The top left panel shows boxplots of residuals for all subjects, indicating that the SLTB model has slightly larger residuals with greater variability compared to the ZOIB model. The top right panel displays paired residuals for each subject, where most lines slope downward, suggesting that residuals tend to be smaller under the ZOIB model. The bottom left panel presents boxplots of residuals for subjects with an outcome of 0. In this subset, the SLTB model again shows slightly more dispersed residuals than ZOIB model. The bottom right panel shows paired residuals for this same subset, further highlighting that residuals are generally smaller under the ZOIB model.}
    \label{fig:slt_zoib_combined}
\end{figure}

MSE for the SLTB regression was 0.00117, which is slightly higher than the MSE of 0.00097 for the ZOIB model. Considering that the ZOIB model estimates two parameters for each variable, one for the (0, 1) portion and one for zero-inflation, whereas the SLTB regression model estimates only one parameter per variable, the difference (0.0002) in their mean squared errors (MSE) for ZOIB and SLTB is relatively small. In the analysis of 1,340 subjects, the SLTB regression model showed a better fit compared to the ZOIB model in terms of squared residuals for 683 subjects, which is 50.97 \% of subjects. 
Among 52 subjects with a response value of 0 for the percentage variable, the SLTB regression model provided a closer approximation to the observed values. The MSE for the SLTB model was 0.00289, which is lower than the ZOIB model’s MSE of 0.00305, indicating smaller absolute residuals for 20 subjects.

 The distribution of residuals for the SLTB regression and ZOIB model is shown in Figure \ref{fig:slt_zoib_combined}. The residual plots suggest that the ZOIB model tends to produce slightly smaller and residuals with lower variability overall at the cost of added complexity from using a separate model for boundary values. However, the SLTB model still provides competitive performance, particularly among subjects with an outcome of 0, where residuals remain close to 0 and show only modest differences compared to ZOIB. Paired residual plots indicate that although ZOIB often yields lower residuals, the differences are relatively small for many subjects. This suggests that despite the added complexity of the ZOIB model, its advantage over SLTB regression in predictive accuracy is modest.

\subsection{SLTB implementation of Bayesian hierachical nonlinear models}
\label{sec: Bayesian nonlinear}

Delay discounting is a behavioral economic phenomena wherein the subjective value of a reward (or outcome) decreases as the delay to receiving that reward increases. A ubiquitous and familiar outcome is the value of money. The present value of \$100 available immediately is \$100 but the present value of \$100 \emph{delayed} in one year may be \emph{discounted} by a certain rate to yield a present value less than \$100. For example, An individual may prefer \$50 today instead of \$100 in six months. In delay discounting research, present values are compared against future values across a range of delays (e.g., one day, six months, one year, five years) and a nonlinear model is fit to these data. Thus, the subjective value \( V \) of a delayed reward is modeled as:

\begin{equation}
    V(D) = \frac{A}{1 + kD} = \frac{A}{1 + exp(\psi)D}
    \label{eq:mazur}
\end{equation}

where \( V(D) \) is the discounted (subjective) value of the reward at delay \( D \), \( A \) is the nominal (objective) amount of the delayed reward, \( D \) is the delay to reward receipt (in consistent time units), and \( k \) is the individual-specific discounting rate parameter (\( k > 0 \)) \citep{mazur2013adjusting}. Delay discounting rate has been shown to be associated with a variety of problematic behaviors including excessive drinking, smoking, gambling, addiction, risky sexual behavior, and others, see e.g.,  \cite{bickel2012excessive}.

The parameter \( k \) is often modeled as $\psi=\text{log}{(k)}$ where $\text{log}(\cdot)$ is the natural log function, and subjective values are commonly converted between 0 (no value) and 1 (the full undiscounted amount). Data for this analysis were indifference points between immediate and delayed job opportunities, and were collected with an algorithm that allowed for indifference points of exactly 0 or 1. Thus, nonlinear beta regression approaches that accommodate boundary values are of interest. %In the delay discounting literature, examining individual's responses for systematicity is typical. This systematicity is described by relatively few reversals (values are assumed to decrease consistently across delays, not fluctuate up and down) and an overall decrease in the subject value at the farthest delay compared to the closest delay. Data used here met both criteria for systematicity. 

Based on the estimated values $\hat{\psi}$ and $\hat{\phi}$, estimated using the maximum likelihood method from the SLTB regression model, we simulated $\psi^{sim}$ and $\ln\phi^{sim}$ assuming the normal distribution as the true underlying distribution for these terms. This follows our preliminary analyses of these data as described in \cite{kim2024thinking}. Using these simulated values of $\psi^{sim}$ and $\ln\phi^{sim}$, we generated indifference point data from the beta distribution. We exclusively simulated data from the beta distribution, as data simulated from a normal distribution would not be valid for discounting data which should be bounded in [0, 1]. 

\[
g(\mu_{ij}^{sim})=\mu_{ij}^{sim} = E(y_{ij}) = \frac{1}{1 + \exp(\psi_i^{sim}) D_j}
\]
\[
y_{ij}^{sim} \mid \psi_i^{sim}, \phi_i^{sim} \sim \text{Beta}(\mu_{ij}^{sim} \cdot \phi_i^{sim}, (1 - \mu_{ij}^{sim}) \cdot \phi_i^{sim})
\]

where $y_{ij}$ is $i$-th subject's $j$th indifference point and $D_j$ is $j$th delay.  
 
A Metropolis-Hastings algorithm was constructed to fit the model, utilizing the following prior distributions:

\[
\psi_i \mid \mu_{\psi}, \sigma^2_{\psi} \sim N(\mu_{\psi}, \sigma^2_{\psi})
\]
\[
\mu_{\psi} \sim N(\mu_{\psi_0}, \Lambda_{\psi_0}^2), \quad \sigma^2_{\psi} \sim IG\left(\frac{a_1}{2}, \frac{b_1}{2}\right)
\]
\[
\text{ln} \ \phi_i \mid \mu_{\phi}, \sigma^2_{\phi} \sim N(\mu_{\phi}, \sigma^2_{\phi})
\]
\[
\mu_{\phi} \sim N(\mu_{\phi_0}, \Lambda_{\phi_0}^2), \quad \sigma^2_{\phi} \sim IG\left(\frac{a_2}{2}, \frac{b_2}{2}\right)
\]

where \( {N}(\cdot, \cdot) \) denotes the normal distribution and \( {IG}(\cdot, \cdot) \) denotes the inverse-gamma distribution. The hyperparameters are chosen to represent weak prior information:

\begin{align*}
    \mu_{\psi_0} = -1, \quad \Lambda^2_{\psi_0} = 10^2, \quad 
    a_1 = 1, \quad b_1 = \frac{1}{10}, \\
    \mu_{\phi_0} = 1, \quad \Lambda^2_{\phi_0} = 10^2, \quad 
    a_2 = 1, \quad b_2 = \frac{1}{10}.
\end{align*}

The Metropolis-Hasting algorithm was initialized using the following values:

\begin{align*}
   \psi^{(0)} &\sim \mathcal{N}(\bar{\psi}, \operatorname{sd}(\psi)), \quad \sigma^{2 \ (0)}_{\psi} = 10^2, \\
   \ln \phi^{(0)} &\sim \mathcal{N}(\overline{\ln \phi}, \operatorname{sd}(\ln \phi)), \quad \sigma^{2 \ (0)}_{\phi} = 10^2.
\end{align*}

where \( \bar{\psi} \) and \( \overline{\ln \phi} \) denote the empirical means, and \( \operatorname{sd}(\psi) \) and \( \operatorname{sd}(\ln \phi) \) represent the empirical standard deviations, calculated from the observed values. 
These initial values are drawn from the empirical distribution of the data, ensuring that the chains begin in regions of high posterior density and reducing the likelihood of poor convergence.

The Metropolis-Hastings algorithm was employed within a mixed-effects model framework, incorporating separate random effects for both $\psi_i$ and $\phi_i$ at the subject level. In the absence of a readily-implemented beta regression-based comparison method for Bayesian hierarchical nonlinear models, we used a model with a similar structure but assuming a normal density. Full details of the normal density model and sampling algorithm can be found in the Online Supplement. Presuming normally distributed residuals for delay discounting data has been a fairly commonly used strategy in the analysis of delay discounting data to date.

For the parameter $\mu_{\psi}$ that quantifies the average log rate of delay discounting, both hierarchical models produced very similar estimates. We note that  trace plots indicate that all algorithms converged successfully.  The normal hierarchical model yielded a posterior mean of -4.93 with a 95\% credible interval of [-5.14, -4.71], while the SLTB hierarchical model produced a posterior mean of -4.87 with a 95\% credible interval of [-5.15, -4.59]. Notably, the posterior mean estimate from the SLTB hierarchical model is closer to the true value of $\mu_{\psi} = -4.87$.

For the parameter $\sigma^2_{\psi}$, the normal hierarchical model resulted in a posterior mean of 1.40 with a 95\% credible interval of [1.05, 1.84], whereas the SLTB hierarchical model provided a slightly higher mean estimate of 2.34 with a wider interval of [1.77, 3.08]. Likewise, the posterior mean estimate from the SLTB hierarchical model is closer to the true value of $\sigma^2_{\psi} = 2.48$.

Overall, both hierarchical models demonstrate reasonable performance in recovering the posterior distribution of the parameters. However, SLTB hierarchical model appears to produce estimates that are closer to the true parameter value.

\begin{table}[H]
\centering
\begin{tabular}{lccc}
\toprule
Model & Lower Bound & Upper Bound & Mean\\
\midrule
SLTB hierarchical model  & -5.15 & -4.59  & -4.87 \\
Normal hierarchical model  & -5.14 & -4.71  & -4.93 \\
\bottomrule
\end{tabular}
\caption{95\% credible intervals for the parameter $\mu_{\psi}$ (truth set to -4.87) using the normal hierarchical model and the SLTB hierarchical model.}
\end{table}

\begin{table}[H]
\centering
\begin{tabular}{lccc}
\toprule
Model & Lower Bound & Upper Bound & Mean\\
\midrule
SLTB hierarchical model  &  1.77 & 3.08  & 2.34 \\
Normal hierarchical model &  1.05 &  1.84 & 1.40 \\
\bottomrule
\end{tabular}
\caption{95\% credible intervals for the parameter $\sigma^2_{\psi}$ (truth set to 2.48) using the normal hierarchical model and the SLTB hierarchical model.}
\end{table}

\section{Discussion}
\label{sec: Discussion}

In this paper, we introduced the SLTB regression model, an extension of the standard beta regression, which accommodates boundary values at 0 and 1. By employing a scale-location transformation and subsequent truncation, the SLTB model overcomes the limitations of the standard beta regression, which cannot handle exact boundary values due to likelihood function constraints. The scale-location transformation expands the support beyond (0,1) to [0, 1], ensuring positive and finite probability distributions even when truncation is applied. The SLTB method works by exploiting the little-recognized fact that truncated densities have positive and finite mass on their boundary values. Therefore, the SLTB regression effectively preserves the fundamental properties of the beta distribution while allowing for positive probability mass at boundary values. Further, the interpretation of SLTB regression parameters is the same as standard beta regression, while mixture model-based approaches have different and potentially less familiar interpretations. The SLTB regression approach uses a smaller number of parameters (albeit with two tuning parameters) compared to mixture model-based approaches. This makes the SLT method more parsimonious overall.

We have shown the flexibility and ease of implementation of the SLTB approach through empirical and simulation studies when boundary values are observed.  Performance comparisons show that the the SLTB regression achieves nearly identical mean squared error (MSE) are comparable to other existing models. Moreover, the results indicate that SLTB regression performs comparably to both XBX regression \citep{kosmidis2025extended} and ZOIB regression \citep{liu2015zoib} despite its relative simplicity. SLTB performs quite well when an available predictor can perfectly separate data in the mixture component responsible for characterizing boundary values, which is analogous to the separation issues in binary regression. In our linear-framework simulation study, ZOIB performed less favorably than SLTB and XBX due to such a separation problem.  Under such conditions, ZOIB is disadvantaged relative to competing approaches. When boundary outcomes are more evenly distributed across groups, as in the \texttt{AlcoholUse} dataset included in the \texttt{ZOIB} package, ZOIB performs as intended.

A key advantage of SLTB is that it does not require explicitly modeling boundary values as separate components, which simplifies estimation and may increase ease of use in practice. In this sense, SLTB is more parsimonious, relying on fewer parameters while still achieving performance comparable to more complex mixture-based approaches. Furthermore, the parameters in SLTB retain the familiar interpretation of standard beta regression, whereas mixture-based models yield coefficients that apply only to interior or boundary observations. This similarity in interpretation may make SLTB preferable in many applications. Finally, SLTB can also serve as a practical alternative to XBX showing computational efficiency and comparable accuracy. 

%\textcolor{red}{In the $N=1,000$ case (see Online Supplement), among the 1,000 Monte Carlo replicates of the simulation study, four of those cases did not converge when using the XBX approach using the betareg software. If such a rare case were encountered in practice, SLTB regression might be a good alternative due to the relative simplicity with which the model is fit. There were no convergence issues for SLTB in the analyses presented in this paper.}
%\textcolor{red}{We deleted the table for this one so it is not in online supplement. Should we delete it or not?}

In addition to accommodating boundary values, the SLTB regression is applicable across a wide range of disciplines where the standard beta regression is commonly used. We have shown that it is relatively simple to employ in both linear and nonlinear frameworks, and its flexibility extends to Bayesian hierarchical modeling scenarios, further enhancing its utility. 

Despite its advantages, the SLTB regression has some limitations. Although SLTB regression successfully models boundary values, further investigation is needed to assess its performance across varying proportions of boundary cases. Additionally, while the scale-location transformation retains the fundamental characteristics of the beta distribution, small deviations may occur depending on choices of scale and location values, which should be carefully considered. While users have the flexibility to adjust scale ($s$) and location ($l$) values, it is important to note that the performance of the SLTB regression model can vary depending on the choice of $s$ and $l$. Therefore, users should carefully evaluate these values based on their specific data and modeling needs.

The Scale-Location-Truncated beta regression model offers a valuable extension to standard beta regression by effectively accommodating boundary values at 0 and 1. Implementation is straightforward: the user simply replaces the standard beta density with the SLTB density, and proceeds with standard model fitting via maximum likelihood or Bayesian approaches. Our study demonstrates that the SLTB regression performs comparably to existing models while offering greater flexibility in handling bounded data. By allowing for positive and finite probability mass at the boundaries, this approach handles long-standing limitations in the standard beta regression modeling.

\bibliographystyle{abbrvnat}
\bibliography{reference}
\newpage
\appendix
\setcounter{section}{0}
\renewcommand{\thesection}{\arabic{section}}

\section*{Online Supplement}
\label{sec: Supplement}

\section{Bayesian Hierarchical Nonlinear Modeling with Metropolis-Hastings}

We estimated the posterior means and 95\% credible intervals for the parameters \(\mu_{\psi}\) and \(\sigma^2_{\psi}\) using two Metropolis-Hastings (MH) models: one assuming a normal distribution and another assuming the beta distribution for its proposal distribution.

\subsection{SLTB Hierarchical Bayesian Model}

\begin{itemize}
    \item \textbf{For each iteration \(s\) from 1 to \(S\):}
    \begin{itemize}
        \item \textbf{Step 1: Update \(\mu_\psi\) (Group-level mean of $\psi_i$)}
        \begin{enumerate}
            \item Assume:
            \[
                \psi_i \sim \mathcal{N}(\mu_\psi, \sigma^2_\psi), \quad \mu_\psi \sim \mathcal{N}(\mu_{\psi_0}, \Lambda_{\psi_0}^2)
            \]
            \item Update \(\mu_{\psi}\) by sampling from the full conditional distribution:
            \begin{align*}
                \mu_{\psi} \mid \psi, \sigma^2_{\psi} &\sim \mathcal{N}(\mu_{\psi_m}, \Lambda_{\psi_m}) \\
                \Lambda_{\psi_m} &= \left( \frac{I}{\sigma^2_{\psi}} + \frac{1}{\Lambda_{\psi_0}^2} \right)^{-1} \\
                \mu_{\psi_m} &= \left( \sum_{i=1}^{I} \frac{\psi_i}{\sigma^2_{\psi}} + \frac{\mu_{\psi_0}}{\Lambda_{\psi_0}^2} \right) \cdot \Lambda_{\psi_m}
            \end{align*}
        \end{enumerate}

        \item \textbf{Step 2: Update \(\sigma^2_\psi\) (Group-level variance of $\psi_i$)}
        \begin{enumerate}
            \item Assume:
            \[
                \psi_i \sim \mathcal{N}(\mu_\psi, \sigma^2_\psi), \quad \sigma^2_\psi \sim \text{IG}\left( \frac{a_1}{2}, \frac{b_1}{2} \right)
            \]
            \item Update \(\sigma^2_{\psi}\) by sampling from the full conditional distribution:
            \[
                \sigma^2_{\psi} \mid \psi, \mu_{\psi} \sim \text{IG} \left( \frac{I + a_1}{2}, \frac{\sum_{i=1}^{I} (\psi_i - \mu_{\psi})^2 + b_1}{2} \right)
            \]
        \end{enumerate}
        
        \item \textbf{Step 3: Update \(\psi_i\) (Individual-level parameters of $\psi$) using Metropolis-Hastings}
        \begin{enumerate}
            \item For each individual \(i\) from 1 to \(I\):
            \item Propose: Sample
            \[
            \psi_i^* \sim \mathcal{N}(\psi_i^{(s)}, 0.5 \cdot \sigma^2_\psi)
            \]
            \item Compute the acceptance ratio:
            \[
            r = \frac{p(\psi_i^* \mid \mu_\psi, \sigma^2_\psi) p(y_i \mid \psi_i^{(s)}, \phi_i^{(s)})}{p(\psi_i^{(s)} \mid \mu_\psi, \sigma^2_\psi) p(y_i \mid \psi_i^*, \phi_i^{(s)})}
            \]
            \item Sample \( u \sim \text{uniform}(0, 1) \). 
            \item Set \(\psi_i^{(s+1)} = \psi_i^*\) if \( u < r \), and \(\psi_i^{(s+1)} = \psi_i^{(s)}\) if \( u \geq r \).
        \end{enumerate}
        
        \item \textbf{Step 4: Update \(\mu_\phi\) (Group-level mean of \(\ln \phi_i\))}
        \begin{enumerate}
            \item Assume:
            \[
                \ln \phi_i^{(s)} \sim \mathcal{N}(\mu_\phi, \sigma^2_\phi), \quad \mu_\phi \sim \mathcal{N}(\mu_{\phi_0}, \Lambda_{\phi_0}^2)
            \]
            \item Update \(\mu_\phi\) by sampling from the full conditional distribution:
            \begin{align*}
                \mu_\phi &\sim \mathcal{N}(\mu_{\phi_m}, \Lambda_{\phi_m}) \\
                \Lambda_{\phi_m} &= \left( \frac{I}{\sigma^2_\phi} + \frac{1}{\Lambda_{\phi_0}^2} \right)^{-1} \\
                \mu_{\phi_m} &= \left( \frac{\sum_{i=1}^I \ln \phi_i^{(s)}}{\sigma^2_\phi} + \frac{\mu_{\phi_0}}{\Lambda_{\phi_0}^2} \right) \cdot \Lambda_{\phi_m}
            \end{align*}
        \end{enumerate}

        \item \textbf{Step 5: Update \(\sigma^2_\phi\) (Variance of \(\ln \phi_i\))}
        \begin{enumerate}
            \item Assume:
            \[
                \ln \phi_i \sim \mathcal{N}(\mu_\phi, \sigma^2_\phi), \quad \sigma^2_\phi \sim \text{IG}\left( \frac{a_2}{2}, \frac{b_2}{2} \right)
            \]
            \item Update \(\sigma^2_\phi\) by sampling from the full conditional distribution:
            \[
                \sigma^2_\phi \mid \ln \phi, \mu_\phi \sim \text{IG} \left( \frac{I + a_2}{2}, \frac{\sum_{i=1}^{I} (\ln \phi_i - \mu_\phi)^2 + b_2}{2} \right)
            \]
        \end{enumerate}
        
        \item \textbf{Step 6: Update \(\ln \phi_i\) (Individual-level parameters of \(\ln \phi\)) using Metropolis-Hastings}
        \begin{enumerate}
            \item For each individual \(i\) from 1 to \(I\):
            \item Propose: Sample
            \[
            \ln \phi_i^* \sim \mathcal{N}(\ln \phi_i^{(s)}, 0.5 \cdot \sigma^2_\phi)
            \]
            \item Compute the acceptance ratio:
            \[
            r = \frac{p(\ln \phi_i^* \mid \mu_\phi, \sigma^2_\phi) p(y_i \mid \psi_i^{(s)}, \phi_i^{(s)})}{p(\ln \phi_i^{(s)} \mid \mu_\phi, \sigma^2_\phi) p(y_i \mid \psi_i^{(s)}, \phi_i^*)}
            \]
            \item Sample \( u \sim \text{uniform}(0, 1) \). 
            \item Set \(\ln \phi_i^{(s+1)} = \ln \phi_i^*\) if \( u < r \), and \(\ln \phi_i^{(s+1)} = \ln \phi_i^{(s)}\) if \( u \geq r \).
        \end{enumerate}
    \end{itemize}
\end{itemize}

\subsection{Normal Hierarchical Bayesian Model}

\begin{itemize}
    \item \textbf{For each iteration \(s\) from 1 to \(S\):}
    \begin{itemize}
        \item \textbf{Step 1: Update \(\mu_\psi\) (Group-level mean of $\psi_i$)}
        \begin{enumerate}
            \item Assume:
            \[
                \psi_i \sim \mathcal{N}(\mu_\psi, \sigma^2_\psi), \quad \mu_\psi \sim \mathcal{N}(\mu_{\psi_0}, \Lambda_{\psi_0}^2)
            \]
            \item Update \(\mu_{\psi}\) by sampling from the full conditional distribution:
            \begin{align*}
                \mu_{\psi} \mid \psi, \sigma^2_{\psi} &\sim \mathcal{N}(\mu_{\psi_m}, \Lambda_{\psi_m}) \\
                \Lambda_{\psi_m} &= \left( \frac{I}{\sigma^2_{\psi}} + \frac{1}{\Lambda_{\psi_0}^2} \right)^{-1} \\
                \mu_{\psi_m} &= \left( \sum_{i=1}^{I} \frac{\psi_i}{\sigma^2_{\psi}} + \frac{\mu_{\psi_0}}{\Lambda_{\psi_0}^2} \right) \cdot \Lambda_{\psi_m}
            \end{align*}
        \end{enumerate}

        \item \textbf{Step 2: Update \(\sigma^2_\psi\) (Group-level variance of $\psi_i$)}
        \begin{enumerate}
            \item Assume:
            \[
                \psi_i \sim \mathcal{N}(\mu_\psi, \sigma^2_\psi), \quad \sigma^2_\psi \sim \text{IG}\left( \frac{a_1}{2}, \frac{b_1}{2} \right)
            \]
            \item Update \(\sigma^2_{\psi}\) by sampling from the full conditional distribution:
            \[
                \sigma^2_{\psi} \mid \psi, \mu_{\psi} \sim \text{IG} \left( \frac{I + a_1}{2}, \frac{\sum_{i=1}^{I} (\psi_i - \mu_{\psi})^2 + b_1}{2} \right)
            \]
        \end{enumerate}

        \item \textbf{Step 3: Update \(\sigma^2\)}
        \begin{enumerate}
            \item Assume:
            \[
            y_{ij} \sim \mathcal{N}\left( \frac{1}{1 + \exp(\psi_i) \cdot D_j}, \, \sigma^2 \right), \quad \sigma^2 \sim \text{IG}\left( \frac{a_2}{2}, \frac{b_2}{2} \right)
            \]
            \item Update \(\sigma^2\) by sampling from the full conditional distribution:
            \[
            \sigma^2 \sim \text{IG} \left( \frac{I \cdot J + a_2}{2}, \frac{ \sum_{i,j}^{I,J} \left( y_{ij} - \frac{1}{1 + \exp(\psi_i) \cdot D_j} \right)^2 + b_2 }{2} \right)
            \]
            \text{ where I is the number of subjects and J is the number of delay points.}
        \end{enumerate}

        \item \textbf{Step 4: Update \(\psi_i\) (Individual-level parameters of $\psi$) using Metropolis-Hastings}
        \begin{enumerate}
            \item For each individual \(i\) from 1 to \(I\):
            \item Propose: Sample
            \[
            \psi_i^* \sim \mathcal{N}(\psi_i^{(s)}, 0.5 \cdot \sigma^2_\psi)
            \]
            \item Compute the acceptance ratio:
            \[
            r = \frac{p(\psi_i^* \mid \mu_\psi, \sigma^2_\psi) p(y_i \mid \frac{1}{1+\exp(\psi_i^{(s)})}, \sigma^2)}{p(\psi_i^{(s)} \mid \mu_\psi, \sigma^2_\psi) p(y_i \mid \frac{1}{1+\exp(\psi_i^*)}, \sigma^2)}
            \]
            \item Sample \( u \sim \text{uniform}(0, 1) \). Set \(\psi_i^{(s+1)} = \psi_i^*\) if \( u < r \), and \(\psi_i^{(s+1)} = \psi_i^{(s)}\) if \( u \geq r \).
        \end{enumerate}

    \end{itemize}
\end{itemize}

\section{SLTB density convergence to beta density}
\begin{sidewaysfigure}[H]
    \centering
    \includegraphics[width=\textheight]{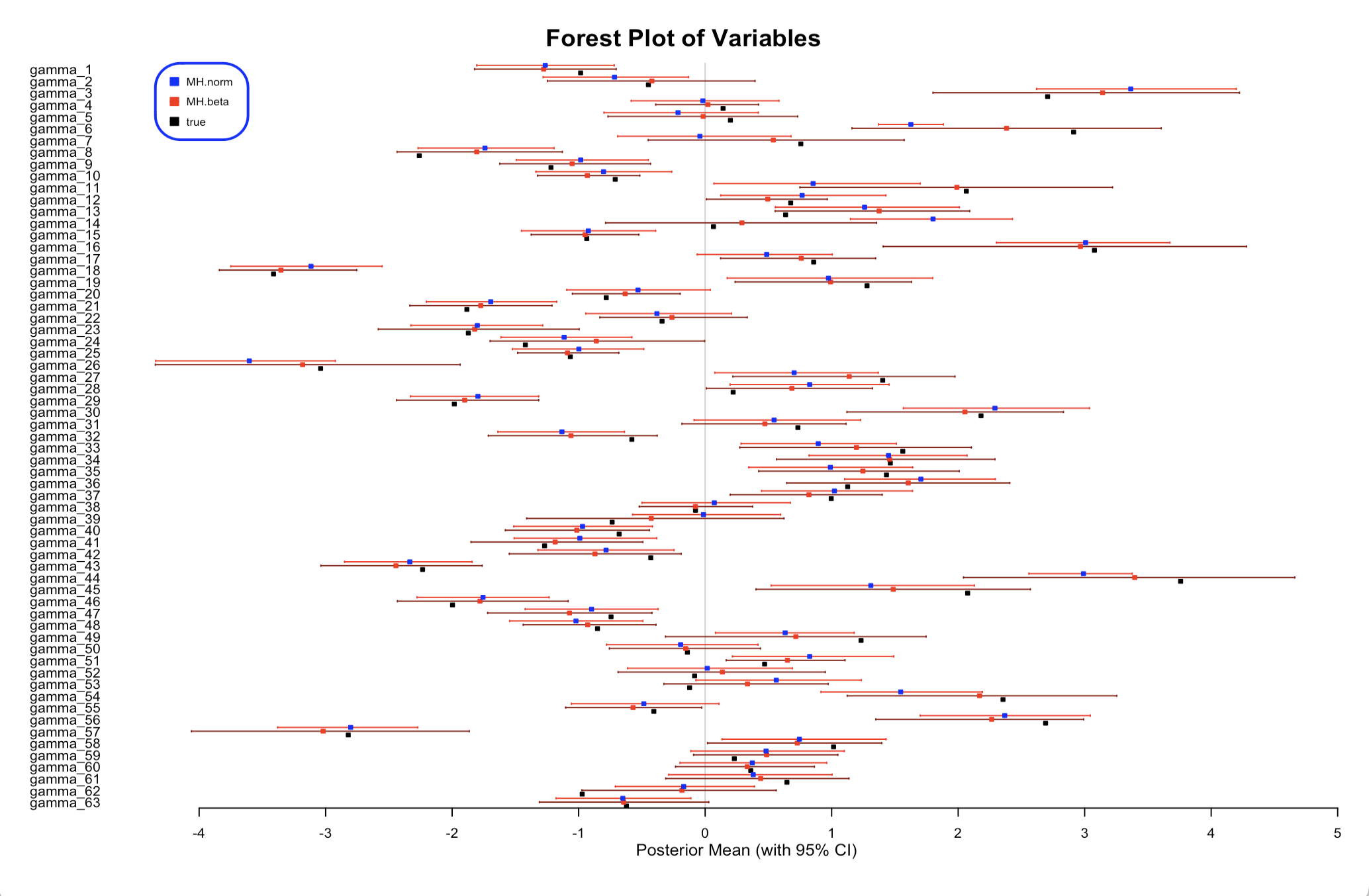}
    \caption{Forest plot comparing the posterior means and 95\% credible intervals of the $\gamma$ parameters under two sampling methods: Metropolis-Hastings with normal proposal (MH.norm) and Metropolis-Hastings with beta proposal (MH.beta). The black dots represent the true parameter values for reference.}
    \label{fig:forestplot1}
\end{sidewaysfigure}

\begin{sidewaysfigure}[H]
    \centering
    \includegraphics[width=\textheight]{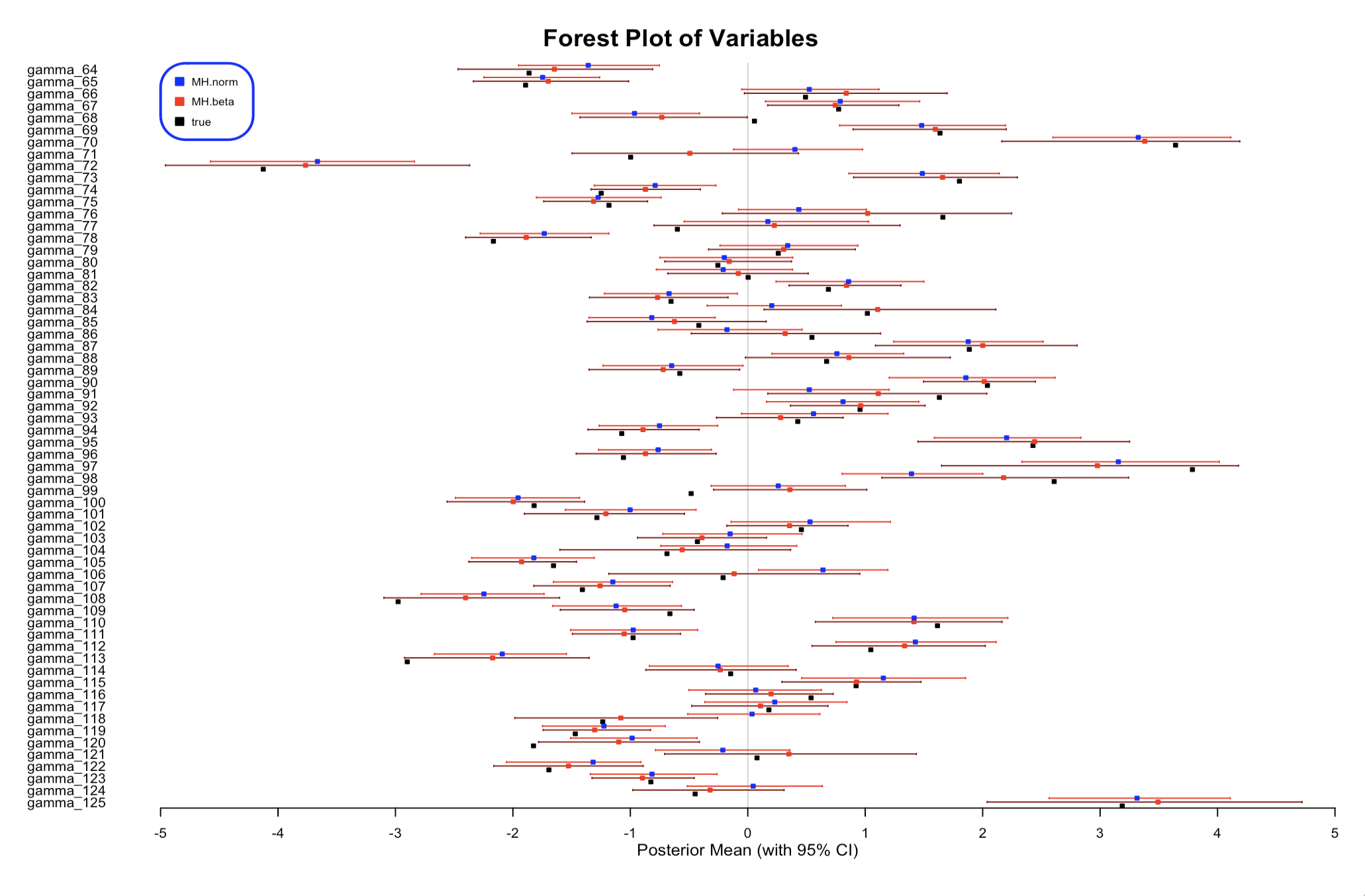}
    \caption{Forest plot comparing the posterior means and 95\% credible intervals of the $\gamma$ parameters under two sampling methods: Metropolis-Hastings with normal proposal (MH.norm) and Metropolis-Hastings with beta proposal (MH.beta). The black dots represent the true parameter values for reference.}
    \label{fig:forestplot2}
\end{sidewaysfigure}

The mean and variance of beta density (\ref{Eq2}) are

\begin{equation}
    \begin{aligned}
        E(Y) &= \left[\frac{\alpha}{\alpha + \beta}\right] \quad & VAR(Y) &=  \left[\frac{\alpha \beta }{(\alpha + \beta)^2(\alpha +\beta+1)}\right]
    \end{aligned}
    \label{beta mean var}
\end{equation}

The difference of mean and variance between SLTB density (\ref{eq: SLTB density}) and beta density (\ref{Eq2}) is 

\begin{equation*}
    \begin{aligned}
        E(G) - E(Y) &= (s-1)\left[\frac{\alpha}{\alpha + \beta}\right] - s \cdot l \\
        \text{Var}(G) - \text{Var}(Y) &= (s^2 - 1) \left[\frac{\alpha \beta }{(\alpha + \beta)^2(\alpha + \beta + 1)}\right]
    \end{aligned}
\end{equation*}

Given that \(s\) is approximately 1 and \(l\) is near 0, the resulting difference is expected to be minimal. For example, we use $s = 1 + 10^{-8.5}$ and $l = 10^{-9}$ in this paper and using this value, the difference will be 

\begin{equation}
    \begin{aligned}
        E(G) - E(Y) &= 10^{-8.5} \left[\frac{\alpha}{\alpha + \beta}\right] - 10^{-9} \\
        \text{Var}(G) - \text{Var}(Y) &= \left(6.324555 \times 10^{-9}\right) \left[\frac{\alpha \beta}{(\alpha + \beta)^2(\alpha + \beta + 1)}\right]
    \end{aligned}
 \label{diff btw SLTB and beta}
\end{equation}

\textbf{Proof:}

Given,
\begin{equation*}
E(G) - E(Y) = (s - 1)\frac{\alpha}{\alpha + \beta} - s \cdot l,
\end{equation*}
\begin{equation*}
\mathrm{Var}(G) - \mathrm{Var}(Y) = (s^2 - 1)\frac{\alpha\beta}{(\alpha+\beta)^2(\alpha+\beta+1)}.
\end{equation*}

Evaluate the limits:

\textbf{Mean:}
\begin{equation*}
\lim_{s\to 1,\,l\to 0}\left[(s - 1)\frac{\alpha}{\alpha + \beta} - s \cdot l\right] = (1-1)\frac{\alpha}{\alpha + \beta} - (1)\cdot(0)=0.
\end{equation*}

\textbf{Variance:}
\begin{equation*}
\lim_{s\to 1}\left[(s^2 - 1)\frac{\alpha\beta}{(\alpha+\beta)^2(\alpha+\beta+1)}\right] = (1^2-1)\frac{\alpha\beta}{(\alpha+\beta)^2(\alpha+\beta+1)}=0.
\end{equation*}

Thus,
\begin{equation}
\lim_{s\to 1,\,l\to 0}[E(G)-E(Y)]=0,\quad \lim_{s\to 1}[\mathrm{Var}(G)-\mathrm{Var}(Y)]=0.\quad\square
\label{Eq:convergence}
\end{equation}

\newpage

%\subsection{SLTB XBX comparisonf Betareg and SLTB Regression}
\label{sec: linear}
\section{SLT XBX comparison on \textit{ReadingSkills} data - no boundary values}

%We analyzed reading accuracy and non-verbal IQ data from the \textit{ReadingSkills} dataset, available in the \texttt{betareg} R package \citep{cribari2010beta}. The dataset consists of 44 observations, with two versions of reading accuracy (accuracy and accuracy1), dyslexia diagnosis, and non-verbal IQ variables. Reading accuracy is proportional data bounded in [0, 1]. While the maximum possible value of the variable accuracy is restricted to 0.99, accuracy1's maximum possible value is 1. In other words, these are identical data except in cases where accuracy has a 0.99 value, accuracy1 has a 1 value. IQ is a numerical variable indicating intelligence quotient transformed to z-scores. Dyslexia diagnosis is represented as a binary factor indicating whether a child is diagnosed as dyslexic. We compare SLTB regression to the  \texttt{betareg} R package using both accuracy and accuracy1 as outcomes. 

We revisit the \textit{ReadingSkills} data and compare the performance of SLTB with standard beta regression. Note that no reading scores have a boundary value in this analysis so standard beta regression is available. We adopt the following model:

\begin{equation*}
    g(\mu_i) = \text{logit}(\mu_i) = \beta_0 + \beta_1 \cdot \text{Dyslexia}_i + \beta_2 \cdot \text{IQ}_i + \beta_3 \cdot (\text{Dyslexia}_i \times \text{IQ}_i).
    \label{eq:_1}
\end{equation*}

For the accuracy outcome variable that does not contain boundary values of 0 or 1, \texttt{betareg} estimates parameters using standard beta regression via a likelihood-based approach (\ref{likelihood}). In contrast, when boundary values are present, i.e., using accuracy1 as an outcome, \texttt{betareg} employs XBX regression \citep{kosmidis2025extended} as an alternative to handle these values effectively.

Based on the SLTB regression model, SLTB regression implementation used a maximum likelihood-based approach to fit to the model in Equation (\ref{eq:_1}).

\begin{table}[H]
\centering
\begin{tabular}{lrrrr}
  \hline
    \textbf{Variable} & \textbf{Estimate} & \textbf{Std. Error} & \textbf{z value} & \textbf{p-value} \\ 
    \hline
     \multicolumn{5}{l}{\textit{Coefficient from SLTB}} \\
  (Intercept) & 1.3338 & 0.1353 & 9.861 & $<$ 0.001 \\ 
  dyslexia    & -0.9736 & 0.1330 & -7.323 & $<$ 0.001 \\ 
  iq          & 0.1607 & 0.1281 & 1.254 & 0.2098 \\ 
  dyslexia:iq & -0.2185 & 0.1282 & -1.705 & 0.0883 \\ 

  \hline
  \multicolumn{5}{l}{\textit{Coefficient from standard beta regression}} \\
  (Intercept) & 1.3338 & 0.1357 & 9.828 & $<$ 0.001 \\ 
  dyslexia    & -0.9736 & 0.1335 & -7.295 & $<$ 0.001 \\ 
  iq          & 0.1608 & 0.1344 & 1.196 & 0.2320 \\ 
  dyslexia:iq & -0.2186 & 0.1345 & -1.624 & 0.1040 \\ 
  \hline
\end{tabular}
\caption{Comparison of coefficient estimates from the SLTB regression models and standard beta regression models. The table reports parameter estimates, standard errors, z-values, and p-values for both models using a logit link function. The models were fitted to the same dataset, including main effects for dyslexia and IQ as well as their interaction. The SLTB model produced estimates nearly identical to those from the standard beta regression, with minor differences in standard errors and z-values.}
\label{tab:betareg_vs_SLTB}
\end{table}

\begin{figure}[H]
    \centering
    \includegraphics[width=0.8\linewidth]{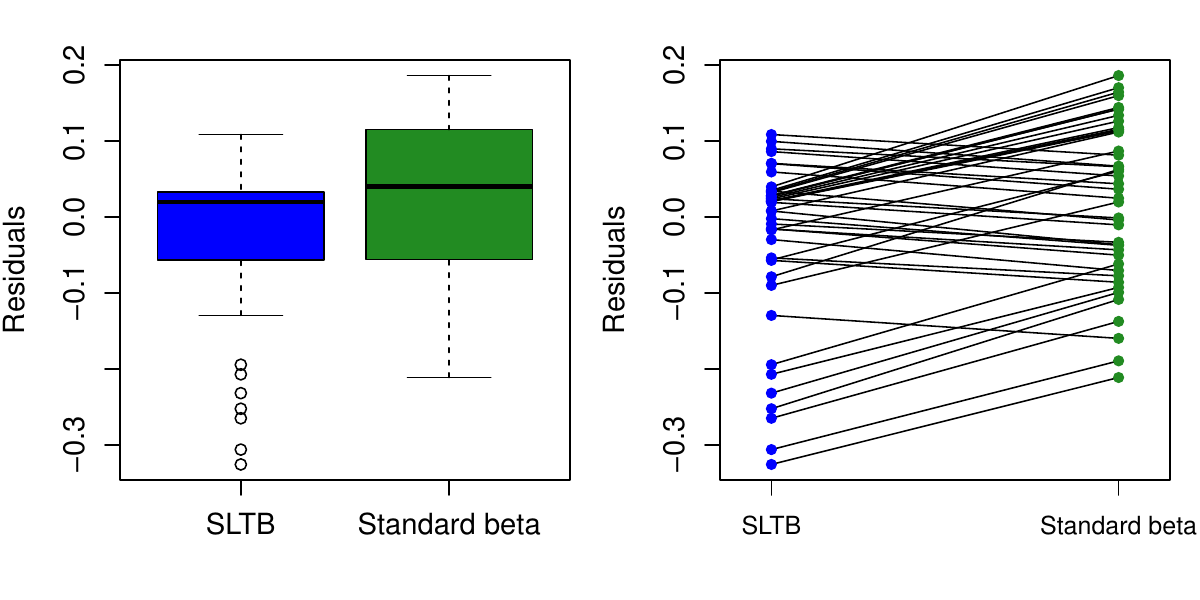}
  \caption{The left panel displays box plots of residuals from the SLTB and standard beta models, showing nearly identical distributions. The right panel illustrates the paired residuals, showing a high degree of correspondence between the two models, with many points exhibiting minimal differences.}
    \label{fig:SLTB_beta}
\end{figure}

When no boundary values are observed in the data, as is the case with accuracy in the Dyslexia set, the SLTB regression model will perform very similarly to standard beta regression owing to the similarity in the densities. To illustrate this, we fit a both the standard beta regression model and the SLTB regression model to the Dyslexia data with accuracy as an outcome. The results are shown in Table \ref{tab:betareg_vs_SLTB}. The SLTB regression model and the standard beta regression model exhibited nearly identical performance in terms of overall model fit and predictive accuracy. Among the 44 observations analyzed, the SLTB model produced lower squared residuals for 21 cases.

The mean squared error (MSE) for the SLTB regression model was 0.00961, closely matching that of the standard beta regression model, which also yielded an MSE of 0.00961. The absolute difference in MSE between the two models was negligible (3.23 $\times$ $10^{-8}$), indicating highly similar overall predictive performance. The distribution of residuals for the SLTB and standard beta regression models is shown in Figure~\ref{fig:slt_process}, highlighting their virtually indistinguishable distributions. These results in the absence of boundary values confirm that SLTB regression performs almost exactly the same as standard beta regression. No surprise, since the densities are extremely similar as shown in Figure \ref{fig:slt_density} of the paper.

\end{document}